# Getting the phase consistent: The importance of phase description in balanced steady-state free precession MRI of multicompartment systems

Nils M.J. Plähn[1,2,3], Simone Poli[2,3,4], Eva S. Peper[1,2], Berk C. Açikgöz[1,2,3], Roland Kreis[2,4,5], Carl Ganter[6], Jessica A.M. Bastiaansen[1,2]

[1] Department of Diagnostic, Interventional and Pediatric Radiology (DIPR), Inselspital, Bern University Hospital, University of Bern, Switzerland

[2] Translation Imaging Center (TIC), Swiss Institute for Translational and Entrepreneurial Medicine, Bern, Switzerland

[3]Graduate School for Cellular and Biomedical Sciences, University of Bern, Bern, Switzerland

[4]MR Methodology, Department for Diagnostic and Interventional Neuroradiology, University of Bern, Bern, Switzerland

[5]Department for Biomedical Research, University of Bern, Bern, Switzerland

[6]School of Medicine, Department of Diagnostic and Interventional Radiology, Klinikum rechts der Isar der TUM, Technical University of Munich, Munich, Germany

**Corresponding author**: Jessica A.M. Bastiaansen, Ph.D.
**Email**: jbastiaansen.mri@gmail.com
**Address**: Laboratory for Quantitative MR Imaging Sciences, Department of Diagnostic, Interventional and Pediatric Radiology (DIPR), Inselspital, Bern University Hospital, University of Bern, Switzerland. Freiburgstrasse 3, 3010 Bern, Switzerland

## Abstract

**Purpose:** Determine the correct mathematical phase description for balanced steady-state free precession (bSSFP) signals in multicompartment systems.

**Theory and Methods:** Based on published bSSFP signal models, two distinct phase descriptions can be formulated: one predicting the presence and the other predicting the absence of destructive interference effects in multicompartment systems. Numerical simulations of bSSFP signals of water and acetone were performed to evaluate the predictions of these two distinct phase descriptions. For experimental validation, bSSFP profiles were measured at 3T using phase-cycled bSSFP acquisitions performed in a phantom containing mixtures of water and acetone, which replicates a system with two signal components. Localized single voxel MRS was performed at 7T to determine the relative chemical-shift of the acetone-water mixtures.

**Results:** Based on the choice of phase description, the simulated bSSFP profiles of water-acetone mixtures varied significantly, either displaying or lacking destructive interference effects, as predicted theoretically. In phantom experiments, destructive interference was consistently observed in the measured bSSFP profiles of water-acetone mixtures, an observation which excludes the phase description that predicts an absence of destructive interference. The connection between the choice of phase description and predicted observation enables an unambiguous experimental identification of the correct phase description for multicompartment bSSFP profiles, which is consistent with Bloch equations.

**Conclusion:** The study emphasizes that consistent phase descriptions are crucial for accurately describing multi-compartment bSSFP signals, as incorrect phase descriptions result in erroneous predictions.

# 1. Introduction

Balanced steady-state free precession (bSSFP) MRI sequences provide images with a high signal-to-noise ratio (SNR) and $T_2/T_1$ contrast[1]. Phase-cycled bSSFP, which refers to multiple MRI acquisitions with different linear phase increments of the radiofrequency (RF) excitation pulse, can be used for banding artifact removal[2–8], the quantification of the $T_1$ and $T_2$ relaxation times[9–13], and the quantification of fat fraction within a voxel[14]. The complex signal as a function of the RF phase increment is a bSSFP profile. For banding artefact removal techniques such as geometric solution[8], as well as for multi-parameter quantification[9–13], the signal phase of the bSSFP profile is a crucial source of information[8–10].

Besides previous studies on water-fat fraction mapping studies[15], most quantitative mapping methods using bSSFP profiles[9–13] assume that a voxel has a single compartment with a unique resonance frequency. In these scenarios, the bSSFP profile appears elliptical on a complex plane and symmetric when plotted against the RF phase increment.

The complex bSSFP profile was described for single compartment systems[2,9–12,16–20], in which the profile has a symmetric and elliptic shape. However, asymmetries in the bSSFP profiles were observed in the presence of gray matter, white matter and muscle[9,21,22], and in voxels containing multiple resonance frequencies such as water and fat[14]. Descriptions for bSSFP profiles in multi-compartments were reported[11] but not thoroughly investigated nor validated for asymmetric signal behavior.

Because certain elements of bSSFP signal phase descriptions may affect how two single component profiles are superimposed, the study aimed to determine the correct mathematical phase description for multi-compartment bSSFP profiles. An intuitive derivation and experimental validation are provided using water acetone mixtures. Acetone was selected over fat due to its improved solubility, and to mimic a scenario with two distinct signal components.

## 2. Theory

In bSSFP, for single component systems, a near zero signal is observed when the RF phase increment matches the phase accumulated between two successive RF excitation pulses due to the local off-resonance frequency[1]. This near-zero signal is responsible for the commonly observed banding artefacts in bSSFP. In multicompartment systems, asymmetric profiles were suggested to originate from convolution between the local frequency distribution and the respective bSSFP profiles[21,22]. Therefore, the final combined profile is an integration of all off-resonant profiles within a voxel, which can be understood using the superposition principle. It can be argued that a near-zero signal obtained with a specific RF phase increment might be the consequence of superposition rather than off-resonance, arising from destructive interference of signals from different tissue types, such as water and fat, within the same imaging voxel. This section aims to describe the destructive interference behavior of bSSFP in multicompartment systems with distinct signal components.

### 2.1. Phase-cycled bSSFP signal model

The signal equation at time $t = 0^+$ directly after the RF excitation pulse of a bSSFP sequence can be written[2,18,20] in the rotating coordinate frame as

$$m_+(0^+, \varphi) = e^{i\phi} \frac{a}{b + c\cos(\vartheta - \varphi)} \left(1 - E_2 e^{-i(\vartheta - \varphi)}\right), \quad [1]$$

with $\varphi$ the RF phase increment and $\vartheta$ the *accumulated phase* as defined in[18], which is a phase directly proportional to local off-resonance i.e., chemical shift $\delta_{cs}$ and magnetic field inhomogeneity $\Delta B_0$. Unless specified, right-handed coordinate systems are used with

$$m_+ = M_x + i \cdot M_y \qquad [2].$$

The constants of Eq. [1] are defined[2,18,20] as $a = M_0(1 - E_1)\sin(\alpha)$, $b = 1 - E_1 E_2^2 + (E_2^2 - E_1)\cos(\alpha)$ and $c = 2(E_1 - 1)E_2 \cos^2\left(\frac{\alpha}{2}\right)$. The (usually unknown) phase factor $e^{i\phi}$ depends on combined effects, such as the chosen RF rotation axis (in the rotating frame), local coil phases, receiver offsets or postprocessing (coil combination, phase regularization). As it neither depends on $\vartheta$ nor $\varphi$, it will be ignored in the following. $E_i = \exp(\frac{-TR}{T_i})$ is the relaxation term describing the longitudinal (T$_1$) and transverse (T$_2$) relaxation time ($i = 1,2$) with repetition time $TR$, RF excitation angle $\alpha$ and the thermal equilibrium magnetization $M_0$. Eq. [2] generates the phase factors of Eq. [1] as $\cos(x) \pm i \cdot \sin(x) = e^{\pm ix}$ which inverts phase signs in a left-handed coordinate system $e^{\pm ix} \to e^{\mp ix}$. Switching between left- and right- handed coordinate systems can be achieved by complex conjugation.

### 2.2. Definition of accumulated phase and phase evolution

Using the accumulated phase definition $\vartheta$, the phase evolution for bSSFP signals at timepoint $t$ can be described as:

$$m_+(t, \varphi) = e^{i\frac{t}{TR}\vartheta} \cdot e^{-\frac{t}{T_2}} \cdot m_+(0^+, \varphi) \qquad [3].$$

To identify the accumulated phase, Eq. [3] can be compared with the solution of the Bloch equation[23]

$$m_+(t, \varphi) = e^{-i\gamma B_0 t} \cdot e^{-\frac{t}{T_2}} \cdot m_+(0^+, \varphi) \qquad [4],$$

with the main magnetic field along the z-direction $\boldsymbol{B}(t) = B_0 \boldsymbol{e}_z$ and for positive gyromagnetic ratios $\gamma$[24]. In rotating coordinate frames, only residual magnetic field inhomogeneity will remain. Hence, the exponent of Eq. [4] can be replaced by $B_0 \rightarrow \Delta B_0$, for isochromatic spins systems. For more than one type of isochromatic spins in a system corresponding to different chemical shifts $\delta_{cs}$, the rotating coordinate frame corresponds to $B_0 \rightarrow (\Delta B_0 + \delta_{cs} B_0)$[17,25]. By comparing Eqs. [3,4] the accumulated phase yields

$$\vartheta = -\gamma(\Delta B_0 + \delta_{cs} B_0) TR \qquad [5].$$

A higher $\delta_{cs}$ value indicates a lower shielding effect of the respective chemical species for their respective nuclei against the $B_0$ field[25].

At this point, a first look at phases and their respective signs can be performed:

- While the minus sign in Eq. [5] follows from the Bloch equations, reconstructed MRI data not necessarily rely on that convention and sometimes ignore it. This merely reflects a choice regarding the handedness of the underlying coordinate system.
- Regardless of the chosen coordinate system, the steady state [1] must depend on the *difference* $\vartheta - \varphi$ of accumulated phase $\vartheta$ and RF phase increment $\varphi$, since changing both by the same amount must not modulate the steady-state amplitude. Getting this wrong may, depending on the situation, also affect quantitative analysis. In this Note, the possible implications of this kind of error will not be further pursued with the assumption that the difference is used.
- At $TE = TR/2$, the phase-related terms in equations [1] and [3] combine (besides small corrections of order $O(1 - E_2)$) to a term proportional to $e^{i\varphi/2} \cdot \sin\frac{\vartheta - \varphi}{2}$. Apart from banding artifacts occurring at $\vartheta \approx \varphi + n \cdot 2\pi$ (independent

of the echo time), this reflects some of the fundamental properties of the bSSFP sequences at $TE = TR/2$:

a) The steady state is $4\pi$-periodic with respect to $\vartheta$.
b) As can also be seen from the multicompartment solution Eq. [7], below, the signal phase, as a function of $\vartheta - \varphi$, is essentially constant ($\approx \varphi/2 + n \cdot \pi$) within any bSSFP plateau (numbered by $n$) between two subsequent banding artifacts. This is also consistent with the well-known spin-echo like behavior of bSSFP[26] for moderate field inhomogeneity.

Properties (a) and (b) no longer hold if the phase evolution in Eq. [3] is replaced by its complex conjugate $e^{-i\frac{t}{TR}\vartheta}$, since at $TE = TR/2$ this would correspond to a multiplication with $e^{-i\frac{\vartheta}{2}}$ instead of $e^{i\frac{\vartheta}{2}}$, such that the phase evolution will become approximately proportional to $e^{-i\vartheta} \cdot e^{i\varphi/2} \cdot \sin\frac{\vartheta-\varphi}{2}$. But for a single isochromat with some fixed (and unknown) $\vartheta$, this erroneous combination of steady state and phase evolution should not be noticeable, since the factor $e^{-i\vartheta}$ would be absorbed in the unknown factor $e^{i\phi}$ in Eq. [1]. As a consequence, quantitative relaxometry, based on several bSSFP acquisitions with different RF phase increments $\varphi_i$, should not be affected by conjugating the phase evolution in Eq. [3], *as long as the tissue is pure*[27].

In the following, it will be investigated whether this still holds, when the tissue of interest is composed of several components with different $\vartheta_i$. To this end, Eq. [3] is multiplied with $e^{i\sigma\frac{t}{TR}\vartheta}$, such that a comparison can be made between the correct ($\sigma = 0$) and the erroneous ($\sigma = -1$) evolution.

### 2.3. Description of destructive interference in multicomponent systems

A system of two components A and B can be described using a single component bSSFP model[2,9–12,16–20] (or Eq.[3]) and the superposition principle[21]:

$$m_{\text{tot}}(t,\varphi) = m_A(t,\varphi) + m_B(t,\varphi) \quad [6]$$

From what was said above, the total signal at $TE = TR/2$ will approximately become

$$m_{\text{tot}}(TE,\varphi) \approx e^{i\phi} \cdot e^{i\varphi/2} \cdot \sum_{j\in\{A,B\}} e^{i\sigma\vartheta_j} \cdot f_j(\vartheta_j - \varphi) \cdot \sin\frac{\vartheta_j - \varphi}{2} \quad [7]$$

where the functions $f_j$ can be assumed to be real and positive, cf. Eq. [1].

Unlike pure tissues considered before, the outcome depends qualitatively on whether the correct ($\sigma = 0$) phase evolution is used or not ($\sigma = -1$):

I. For $\sigma = 0$, the sum in [7] is real and the total phase can only take two opposed values: $\phi + \varphi/2$, if the sum is positive and $\phi + \varphi/2 + \pi$ otherwise. Due to the periodicity of each addend, the sum at any $\varphi$ and $\varphi + 2\pi$ must be of equal magnitude but of opposite sign. Therefore, there must be values $\varphi$, where the signal $m_{\text{tot}}(TE,\varphi)$, becomes zero. Note that this is not caused by banding artifacts associated with each component, but rather due to destructive interference. This cancellation must also be observable in more complex tissue compositions, consisting of three or more components.

II. For $\sigma = -1$, the addends become complex due to the factors $e^{-i\vartheta_j}$, which will usually vary from component to component. Unlike the pure case, these factors can no longer be dragged out of the sum and absorbed in the global phase factor $e^{i\phi}$ and are therefore observable: The sum will decompose into a real and imaginary component, each of which will exhibit destructive interference, but in general not for the same $\varphi$. Therefore, the magnitude of $m_{\text{tot}}(TE,\varphi)$ cannot be expected to vanish for any $\varphi$. Also, the phase of $m_{\text{tot}}(TE,\varphi)$ will show a more complicated behavior than the linear dependence on $\varphi$, which was obtained for $\sigma = 0$.

# 3. Methods

## 3.1. Simulation experiments

bSSFP profiles were simulated using MATLAB R2022a (The MathWorks Inc, Natick, MA) using Eqs. [4,6] with $\sigma = 0$ and $\sigma = -1$ for left and right coordinate handedness. Profiles were simulated for a 36%, 60% and 100% proton density acetone-water fraction, for different $TR = [3.40, 3.62, 4.11, 4.62, 4.84, 5.14]$ms with $TE = \frac{TR}{2}$, $T_1 = 3s$, $T_2 = 2.1s$, for both water and acetone. Additional parameters were $B_0 = 2.89T$, $\Delta B_0 = 0T$, $\delta_{cs,water} = 0ppm$, $\alpha = 35°$ and 38 RF phase increment values $\varphi = \frac{2\pi}{38}(k-1)$ for $1 \leq k \leq 38$. Based on experimental results, the relative chemical shift of acetone was fixed to -2.3ppm and -2.5ppm for the 60% and 36% acetone-water mixtures, respectively. For visual comparison the simulated profiles were rotated using the negative angle of the complex sum of the profiles[10].

$\Delta B_0$ values and $e^{i\phi}$ factors (cf. Eqs. [4,5]) where randomized to check profile shapes for invariances to $\Delta B_0$ values and global phase factors.

## 3.2. MRI experiments in an acetone-water phantom at 3T

The bSSFP profiles were measured at 3T (MAGNETOM Prisma, Siemens Healthineers) using 18 phase-cycled bSSFP acquisitions with $\varphi = \frac{2\pi}{18}(k-1)$ and $1 \leq k \leq 18$, $\alpha = 35°$, $TR = [3.40, 3.62, 4.11, 4.62, 4.84, 5.14]$ms, $TE = \frac{TR}{2}$ and $(1.5mm)^3$ isotropic resolution, in an acetone-water phantom, each contributing with their own distinct signal and phase behavior to the total measured bSSFP signal. The phantom contains 11 vials of 5mL, each with a different acetone-water fraction, immersed in 2% agar. Acetone and water were mixed by recording the weight of the substances, which were then used to calculate proton densities, as described in[28].

Because data were acquired in a left-handed coordinate system[29,30], bSSFP profiles were complex conjugated and then plotted in the (right-handed) complex plane to allow comparison with simulated data. To improve visual comparison, the bSSFP profiles were rotated using the negative angle of their complex sum before plotting.

### 3.3. MRS experiments at 7T

Localized single voxel MRS was performed on the phantom vials containing $f_{PD} = 36\%$ and $f_{PD} = 60\%$ acetone to verify the chemical shifts. For MRS, a STEAM[31] sequence was used at 7T (MAGNETOM Terra, Siemens Healthineers) with a $TR/TE/TM = 5s/20ms/13ms$, 32 averages, a voxel of 10x10x10mm3 , a spectral width of 5000Hz and 4096 data points. Spectra were processed with JMRUI[32] and fitted using AMARES[33] to obtain the chemical shifts.

## 4. Results

### 4.1. Simulation results:

For single components, the elliptical shape of the bSSFP profile is independent of the phase definition of $\sigma = 0$ and $\sigma = -1$ and for randomized $\Delta B_0$ values and global phase factors $e^{i\phi}$ (**Figure 1**). Different $e^{i\phi}$ factors led to a rotation of the profile in the complex plane. Note that this rotation is not visible because the profiles were rotated during post-processing. Changing the $\Delta B_0$ values rotate the sampled points on the elliptical trajectories, while leaving the trajectories themselves unchanged.

Superposition was applied to obtain two-component profiles of water and acetone (**Figure 2**). For TR=3.62ms the phase difference of the water (**Figure 2D**) and acetone (**Figure 2E**) profile was 180°, resulting in a combined profile that is symmetric, appearing as a single component profile.

Two-component simulations demonstrate that different phase definitions ($\sigma = 0$ or $\sigma = -1$) lead to significant different profile shapes (**Figure 3** and **4**), appearing either ribbon shaped, or heart shaped. Complex conjugation leads to mirroring of the profile shape (**Figure 3** and **4,** columns 1 and 2), which implies that a 60% acetone fraction profile with $\sigma = 0$ (**Figure 3M,P**, dashed box) would appear similar to a ~40% acetone fraction profile that is complex conjugated (**Figure 4J**, dotted box).

For the $\sigma = -1$ case, destructive interference was observed for TR values close to 4.84ms (**Figure 4S**), and also for TR=3.62ms (**Figure 3G**). For other TR choices, the asymmetric profiles are appearing heart-shaped centered around origin, without destructive interference (**Figure 3** and **4** third column), confirming theory.

For $\sigma = 0$, destructive interference is present for all TR values (**Figure 3** and **4** column 1 and 2), as expected according to theory.

### 4.2. MRI experiments

In the 100% acetone vial, which mimics a single component system, the bSSFP profiles are symmetric and elliptical (**Figure 1**). In a few of these bSSFP profiles, a gap can be observed within the ellipse, indicating B0-drift (**Figure 1T** and **3T**, red arrow).

In vials containing acetone-water fractions of 36% and 60%, the bSSFP profiles appear ribbon-shaped (**Figure 3** and **4**, last column) and are similar to profiles simulated using the $\sigma = 0$ case. For this case, for every TR, the observed destructive interference matches predictions (**Figure 3** and **4,** first column). In these experimental bSSFP profiles, the absence of destructive interference was never observed, which contradicts with the predicted profiles in the $\sigma = -1$ case.

Interestingly, the bSSFP profiles obtained in the 60% fraction (**Figure 3P**) and the 36% fraction (**Figure 4L**), appear similar, but mirrored from one another. These results show that knowledge of the coordinate handedness is mandatory for a correct evaluation of tissue fractions based on experimental bSSFP profiles (cf. **Figure 3** and **4**, dashed and dotted boxes).

The chemical shift difference of acetone and water, measured with MRS, was -2.32±0.09 ppm for the 60% and -2.48±0.09 ppm for the 36% acetone mixtures (**Figure 5**).

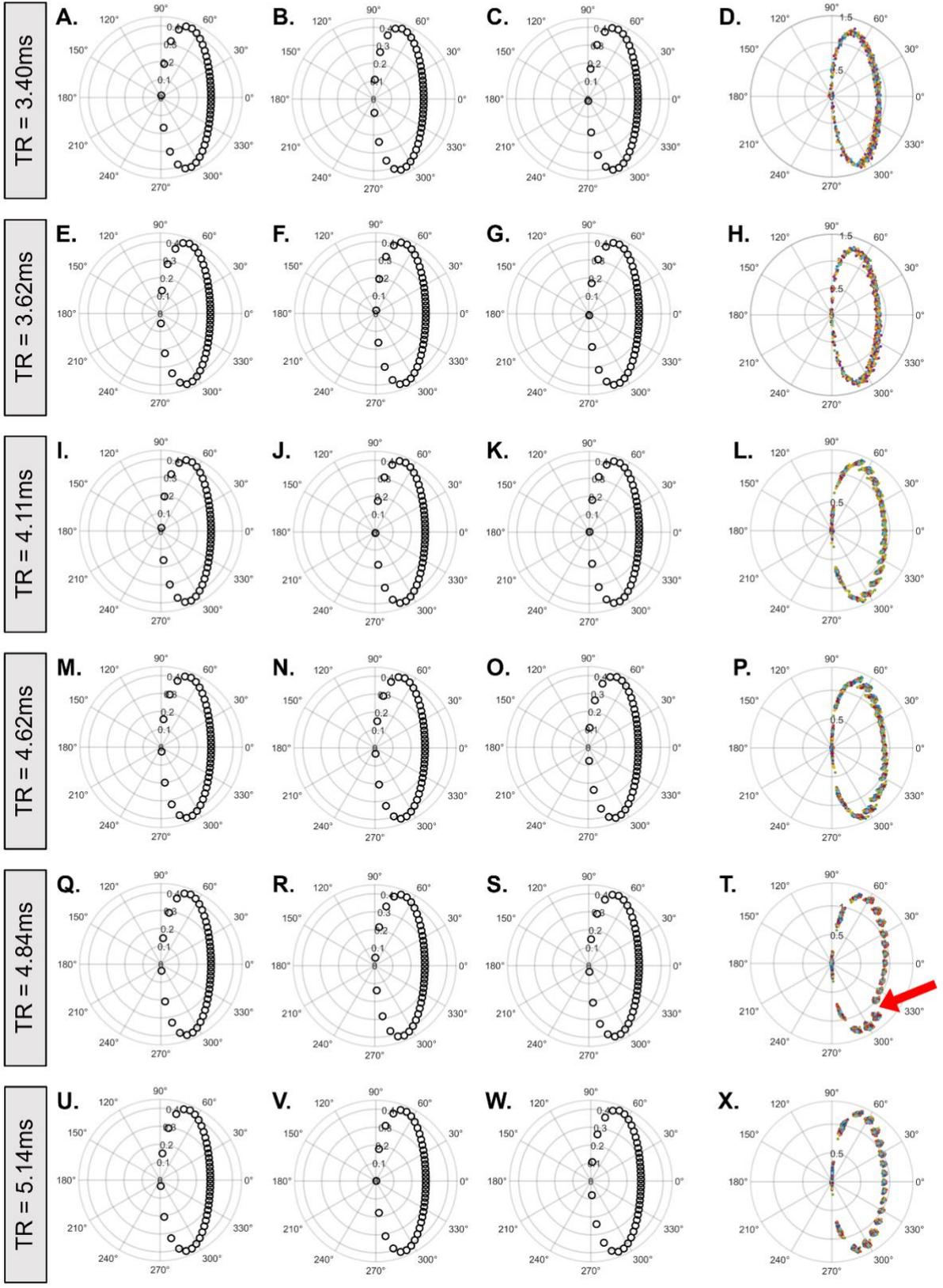

**Figure 1.**

Simulated and experimental bSSFP signal profiles obtained for a 100% acetone solution. All profiles remain elliptical for different repetition times. The red arrow indicates B0 drift effects. Banding is observed at the origin of the complex plots where the signal magnitude is near zero.

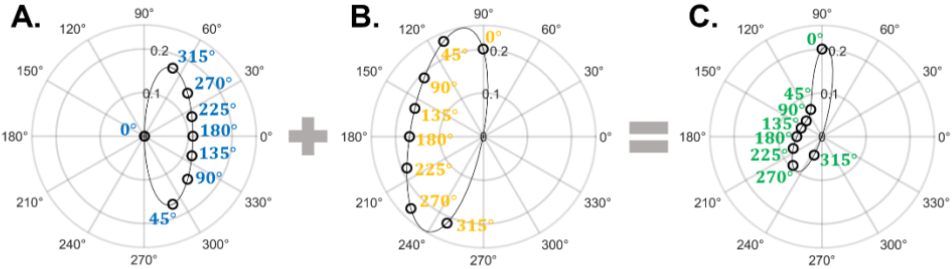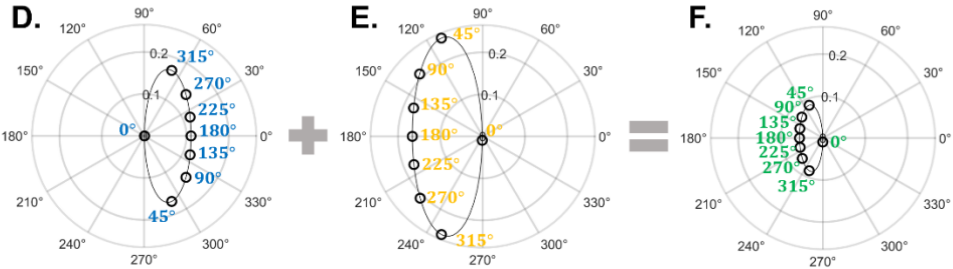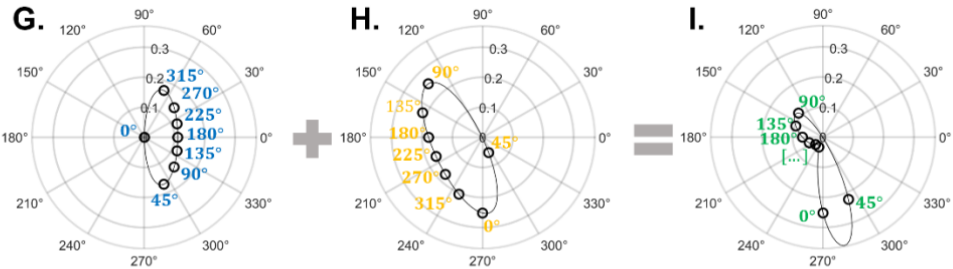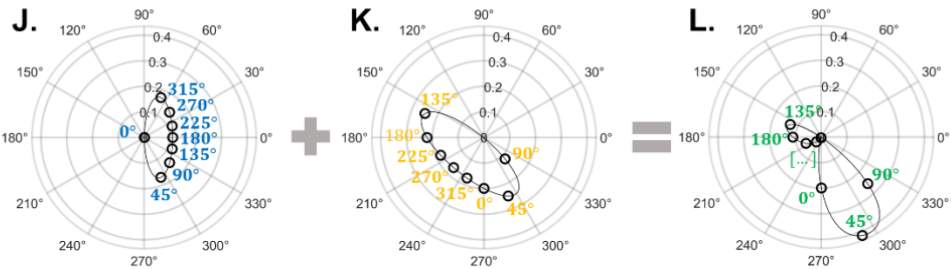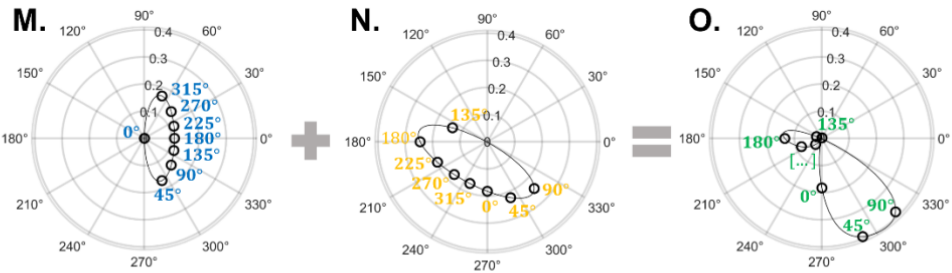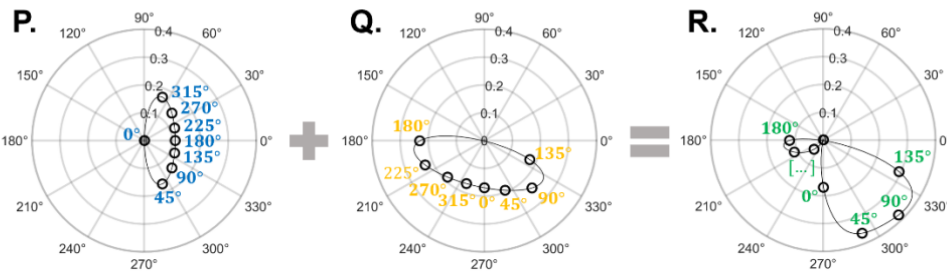

**Figure 2.** Simulated bSSFP profiles for water (blue), acetone (yellow), and a 60% acetone-water mixture (green) using the correct phase sign case of $\sigma = 0$. Simulations were performed for different TR values. The RF phase increments are indicated in the plots in degrees (°). For TR=3.62, the acetone and water profiles are shifted 180 degrees, and the combined profile appears as a single component profile. Banding is observed at the origin of the complex plane of the single component profiles where the signal magnitude is near zero. Destructive interference is observed at the origin of the complex plane of the superimposed profile, where the signal magnitude is near zero.

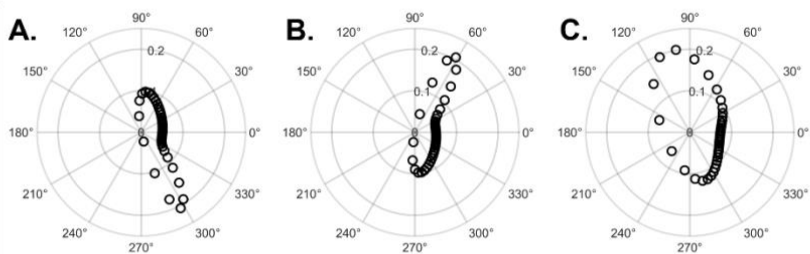
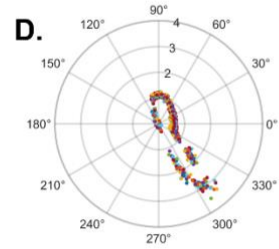
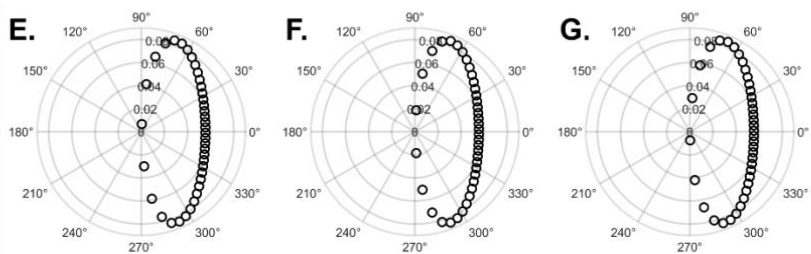
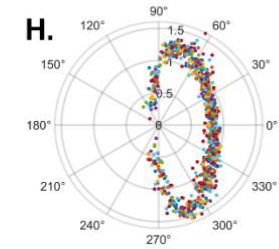
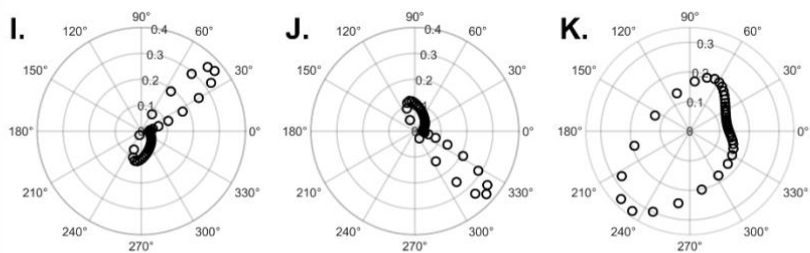
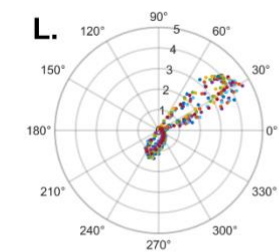
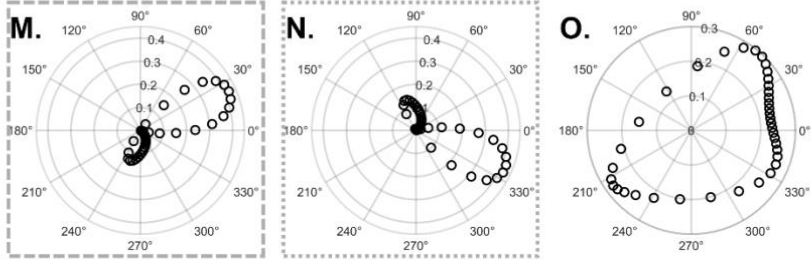
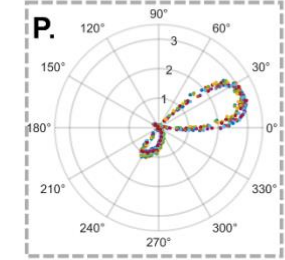
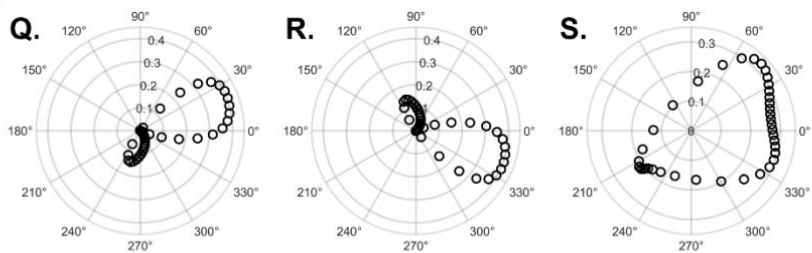
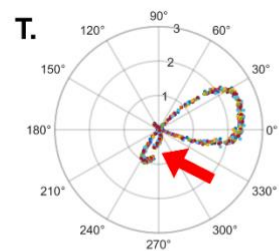
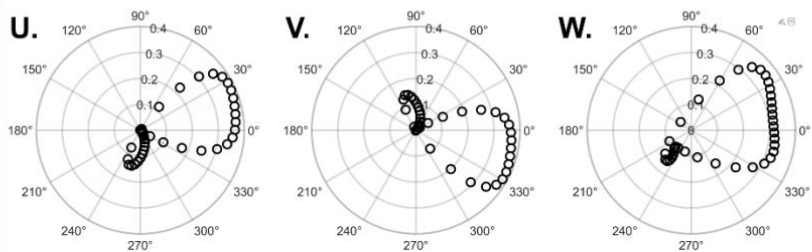
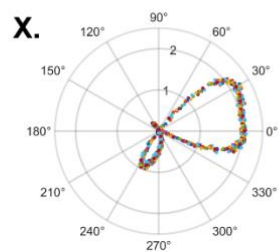

**Figure 3.** Simulated and experimental bSSFP signal profiles obtained for a 60% acetone-water mixture. Simulated profiles appear different depending on the phase description: The $\sigma = 0$ case shows ribbon shaped profiles while the $\sigma = -1$ case shows heart-shaped profiles. Symmetric profiles appear when TR=3.62ms, in this case, the two-component system appears as a single-component system. The dashed and dotted boxes indicate bSSFP profile shapes that appear mirrored for TR=4.62ms. The red arrow in panel T indicates B0 drift effects. Destructive interference is observed at the origin of the complex plane where the signal magnitude is near zero.

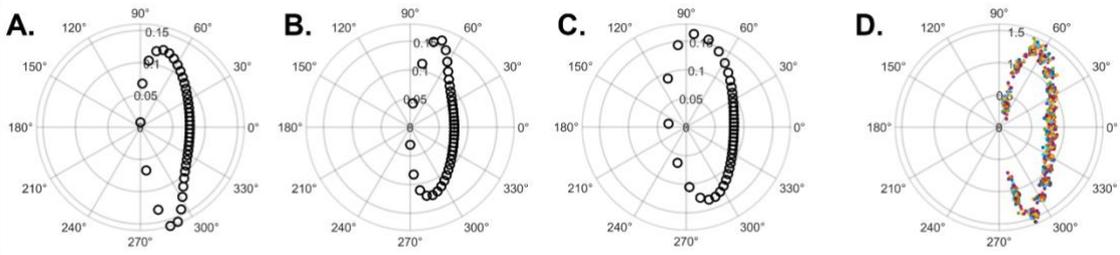
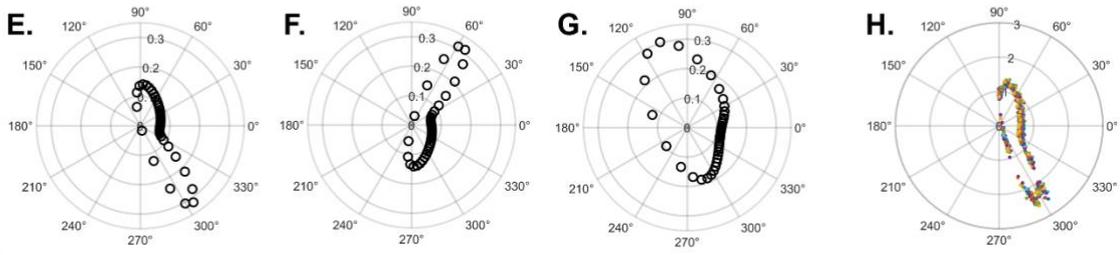
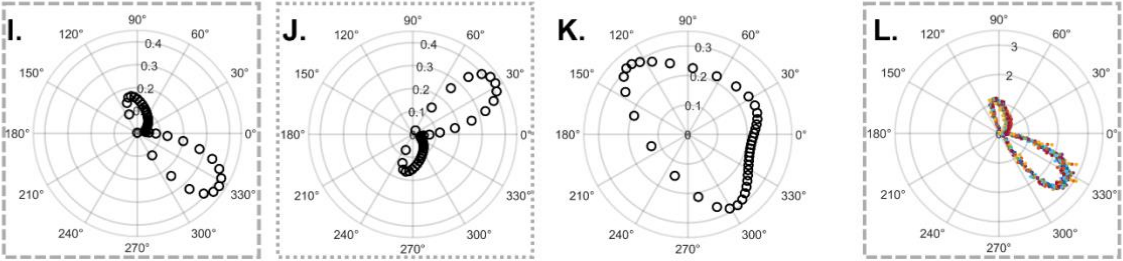
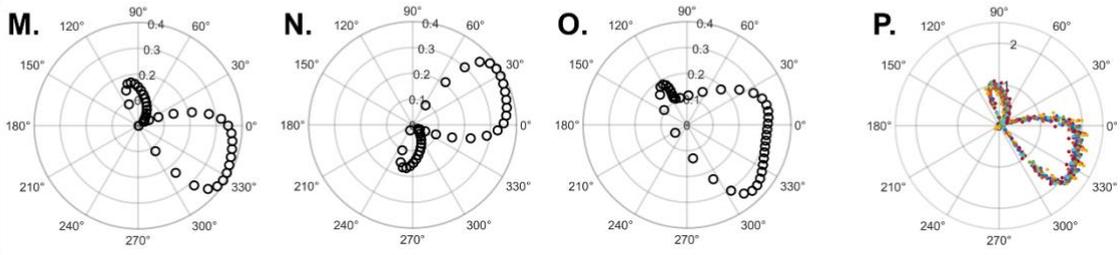
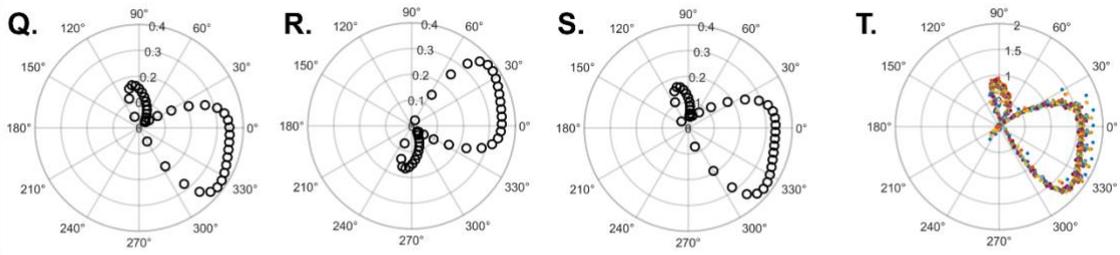
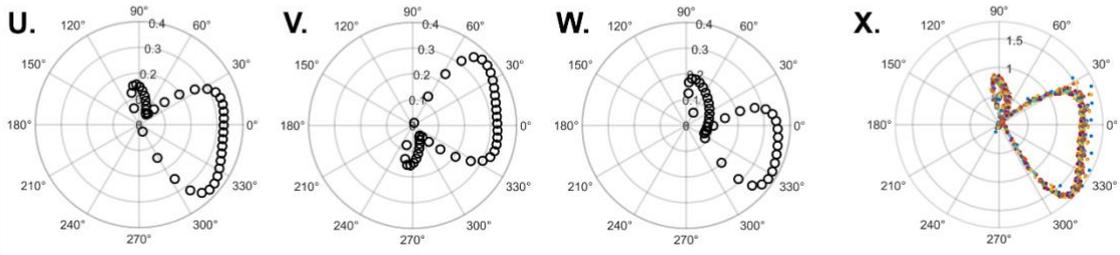

**Figure 4.**

Simulated and experimental bSSFP signal profiles obtained for a 36% acetone-water mixture. Simulated profiles appear different depending on the phase description: The $\sigma = 0$ case shows ribbon shaped profiles while the $\sigma = -1$ case shows heart-shaped profiles. Profiles are nearly symmetric when TR=3.40ms. The dashed and dotted boxes are indicating the mirrored profile shapes. Destructive interference is observed at the origin of the complex plane where the signal magnitude is near zero.

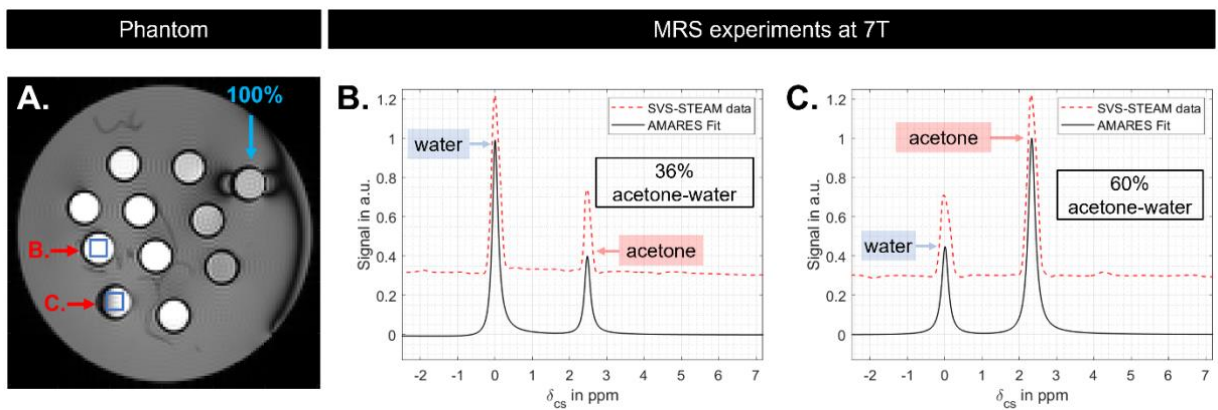

**Figure 5.**

A) Acetone phantom of different vials for different acetone-water mixtures for a RF phase increment of 180°. The blue arrow indicates the 100% acetone vial, while the red arrows indicate image B and C. B&C) Localized MRS performed at 7T in the vials containing a 36% and 60% proton density acetone-water fraction. The relative chemical shift of acetone is concentration dependent and was determined to be -2.48ppm for 36% and -2.32ppm for the 60% proton density acetone fraction relative to the chemical shift of water.

## 5. Discussion

This work demonstrated the importance of correct phase descriptions (referring to $\sigma = 0$ in Eq. [7]) for bSSFP profiles in multicompartment systems, which was shown theoretically and confirmed by simulations and experiments. In contrast to symmetric profiles, which are typically observed in single component systems, an incorrect phase description can lead to erroneous predictions of asymmetric profiles for multi-component and compartment systems, which was tested and validated using acetone-water mixtures. Furthermore, to quantify component fractions correctly, knowledge about coordinate handedness of the data acquisition, which typically depends on the vendor, is crucial.

In single component systems, using either phase description has no effect on the bSSFP profile shapes, because such profiles are not subject to superposition and are thus symmetric. The shape is not affected by the changing coordinate handedness. These observations agree with the different phase descriptions reported in prior work[2,9–12,16–19]. Nevertheless, although the shape of the bSSFP profile remains unaffected, the data points could be arranged clockwise or counterclockwise, which can affect certain T1 and T2 quantification methods[9].

However, for two-compartment systems there is a clear correct ($\sigma = 0$) and incorrect ($\sigma = -1$) way to combine the steady state signals and subsequent phase evolution. The combination of two single-component profiles, using superposition, leads to significantly different bSSFP profile shapes depending on the chosen phase description in Eq. [7]. For the case where $\sigma = 0$, destructive interferences were consistently observed, and for the case $\sigma = -1$, they were mostly absent. Therefore, being inconsistent with the phase leads to wrong bSSFP profile shapes and is therefore likely to cause systematic errors.

In addition, the choice of the coordinate handedness has an effect on the "chirality" of the predicted profile, which is crucial for the correct estimation of tissue fractions. The use of right- or left-handed coordinate systems is not interchangeable because it causes a mirroring of the asymmetric profile and therefore corresponds to a different acetone-water fraction. In essence, this could be viewed as multiplying the underlying chemical shifts with -1, or as a straightforward tissue swap. For example, a complex conjugated profile of 60% acetone-water fraction corresponds to a non-complex conjugated profile of a 40% acetone-water fraction.

In experimental data, the bSSFP profiles of 60% acetone-water appear as single component profiles for TR=3.62ms. This is expected because for certain TR values and chemical shifts, the individual bSSFP profiles of the two components may be shifted by 180 ° or 0 ° relative to each other, leading to two-component profiles which are visually perceived as elliptical single-component profiles. Therefore, it would be possible to estimate the chemical shift difference between water and acetone using bSSFP profile measurements. A rough chemical shift estimation based on bSSFP profiles agreed well with chemical shifts obtained with MRS. The observation that the chemical shift difference between acetone and water depends on the acetone-water fractions suggests that these types of mixtures are not well suited to validate tissue fraction quantification techniques such as those performed with Dixon techniques[34,35], unless verified with spectroscopy.

In summary, working with a consistent phase description ($\sigma = 0$) is crucial to correctly describe multicomponent and multicompartment bSSFP profiles, specifically to avoid possible systematic errors in the analysis of bSSFP profile asymmetries e.g. at higher field strengths[27], for T1 and T2 quantification in gray matter, white matter and in muscles[9,11,21,22], fat-water separation techniques[4,5,36] and fat quantification techniques using phase-cycled bSSFP[14].

## 6. Conclusion

The study demonstrates the vital importance of a consistent and correct treatment of phases in bSSFP signal descriptions of multi-compartment systems. The use of incorrect phase descriptions leads to erroneous bSSFP profiles as well as erroneous quantitative analysis of multi-compartment systems, which has been demonstrated both theoretically and experimentally.

## 7. Acknowledgements

This study was supported by funding received from the Swiss National Science Foundation, grant #PCEFP2_194296.

## 8. Data availability

The acquired data are available for download in the following repository, https://github.com/QIS-MRI/bSSFP-Getting-the-phase-consistent.